\documentclass[final]{aipproc}
\layoutstyle{6x9}
\usepackage{epstopdf}
\usepackage{graphicx}
\usepackage{rotating}
\def\beq{\begin{equation}}
\def\eeq{\end{equation}}
\def\bea{\begin{eqnarray}}
\def\eea{\end{eqnarray}}  
\def\eq#1{{Eq.~(\ref{#1})}}

\newcommand{\Lb}{\left(}
\newcommand{\Rb}{\right)}
\parskip=\itemsep

\def\pom{{I\!\!P}}
\begin{document}
\title{The Interplay Between Data and Theory in Recent Unitarity Models}

\author{Uri Maor}{
address={Department of Particle Physics, School of Physics and Astronomy,
Raymond and Beverly Sackler Faculty of Exact Science,
Tel Aviv University, Tel Aviv 69978, Israel}
\vskip0.2cm
\bf{Talk presented at Diffraction 2008, La Londe-les-Maures, France, September 
2008.} 
}
\begin{abstract}
The role of data analysis in the formation of recent unitarity models 
is discused and evaluated. It is claimed that present support for multi Pomeron 
enhancement beyond the zero order is marginal and should be checked at LHC and 
Auger.
\end{abstract}
\maketitle
\section{Introduction}
\par
The study of s-channel unitarity and quest for its signatures
is a fundamental issue which is instrumental for
estimates of inelastic hard diffraction rates, notably,
diffractive Higgs production at the LHC for which 
the knowledge of the survival probabilities is essential\cite{heralhc}.
Thus, reliable modeling of high energy soft scattering
carries far reaching consequences.
Unitarity models depend not only on
their formulation but critically, also, on their coupled data analysis. 
In this presentation I shall discuss the interplay
between data and theory in recent unitarity models\cite{GLMM,RMK,LKMR,TT}.  
\par
The $b$-space unitarity equation in a diagonal single channel
representation is
\beq \label{unit}
Im\,A^S(s,b)=|A^S(s,b)|^2\,+\,G^{in}(s,b).
\eeq
A general solution of \eq{unit} can be written as    
$A^S(s,b)=i(1-\exp(-\frac{1}{2}\Omega^S(s,b)))$ and  
$G^{in}(s,b)=(1-\exp(-\Omega^S(s,b)))$ with an arbitrary 
opacity $\Omega^S$. 
In eikonal models $\Omega^S$ is real 
and is given by the input, non screened, amplitude.  
In the eikonal approximation the unitarity bound coincides with 
the black bound, $|A^S(s,b)|\leq 1$.  
Note, though, that enforcing unitarity is model dependent. 
The non eikonal model of Ref.\cite{TT} is 
a manifestation of the above, 
in which $|A^S(s,b)|\leq 2$. This  
implies that at high enough energies 
$\sigma_{el}$ approaches $\sigma_{tot}$ logarithmically. 
The LHC predictions of this model are dramatic: $\sigma_{tot}=230 mb$ and   
$\sigma_{el}=150 mb$. The consequent survival probabilities for 
hard diffractive channels are very close to one.  
These bold predictions will soon be checked!   
\section{Eikonal Good-Walker Models}
\par
Updated eikonal models take into account both elastic and diffractive 
re-scatterings of the initial projectiles\cite{heralhc}. 
This is a consequence of the Good-Walker mechanism\cite{GW} in which 
the proton (anti proton) wave function has elastic and low mass diffractive 
components which scatter elastically.
For each of the 3 independent amplitudes we write
\beq \label{UNIT}
Im\,A^S_{i,k}(s,b)=|A^S_{i,k}(s,b)|^2+G^{in}_{i,k}(s,b),
\eeq
in which 
$A_{i,k}^S(s,b)=i(1-\exp(-\frac{1}{2}\Omega_{i,k}^S(s,b)))$ and
$G_{i,k}^{in}(s,b)=(1-\exp(-\Omega_{i,k}^S(s,b)))$.
\newline
$G^{in}_{i,k}$ is the summed probability for
all non GW induced inelastic final states.
\par
The eikonal models I shall discuss\cite{GLMM,RMK,LKMR} 
aim to describe very high energy scattering 
initiated by a Pomeron exchange, where 
$\alpha_{\pom}(t)=1+\Delta_{\pom}+\alpha_{\pom}^{\prime}t$. We get 
$\Omega^S_{i,k}(s,b)\,=\,\nu^S_{i,k}(s)\Gamma^S_{i,k}(s,b,\alpha_{\pom}^{\prime})$, 
where $\nu^S_{i,k}(s)\,=\,g_{i}g_{k}(\frac{s}{s_0})^{\Delta_{\pom}}$ and  
$\Gamma^S_{i,k}$ are the $b$-space profiles which control 
the $t$ dependence of the scattering amplitudes. 
This is external information derived from the 
data analysis of the $t$ dependences. 
\par
Assume a model in which diffraction is exclusively GW.
The output elastic, SD and DD amplitudes are  
linear combinations of $A^S_{i,k}$. 
We have to determine 5 free Pomeron parameters to which we add   
2 parameters needed to construct $\Gamma^S_{i,k}$. The relevant data we 
have in the SPPS-Tevatron energy range is not sufficient to constrain that many 
parameters. Refs.\cite{GLMM,LKMR}  
overcome this difficulty by adding the ISR data to their fits 
which necessitates  the addition of secondary Regge exchanges.  
KMR ignore this problem as they do not fit the data.
The common features of the above GW models are:
\newline
1) They reproduce the elastic sector of their data base 
remarkably well (our $\chi^2/dof=0.87$), 
but fail to reproduce the diffractive sector. 
\newline
2) All have remarkably small $\alpha_{\pom}^{\prime}$. 
GLMM fitted $\alpha^{\prime}_{\pom}=0.012$, LKMR fit is 
$\alpha^{\prime}_{\pom}=0.033$.  
KMR assume $\alpha^{\prime}_{\pom}=0$, while in an earlier KMR 
edition\cite{KMR} $\alpha^{\prime}_{\pom}=0.066$.
\newline
3) Refs.\cite{GLMM} and \cite{LKMR} fit $\Delta_{\pom}=0.12$.  
It is not specified in Ref.\cite{RMK}. Ref.\cite{KMR} fit 
$\Delta_{\pom}=0.10$. 
\section{Enhanced Pomeron Contributions}
\par
GW models short comings are amended once multi Pomeron 
interactions, leading to high mass diffraction, are included. The models 
considered are based on similar basic philosophy but their modeling 
are significantly different.
In GLM the smallness of $\alpha_{\pom}^{\prime}$ leads to the conclusion that the 
soft, seemingly non perturbative, interactions are hard enough 
to enable pQCD calculations. 
Even though KMR assume that $\alpha_{\pom}^{\prime}=0$, 
theirs is a Reggeon Calculus with a partonic interpretation. 
LKMR is a much simpler model where Pomeron enhancement 
is reduced to its zero order triple Pomeron approximation 
for SD. DD is not discussed. 
The inclusion of Pomeron enhancement does not significantly 
change the GW Pomeron parameters with the exception of 
$\Delta_{\pom}$ which becomes larger.
GLMM fitted $\Delta_{\pom}=0.34$, while KMR tuned $\Delta_{\pom}=0.55$. 
LKMR maintain their GW low $\Delta_{\pom}=0.12$.
The high input $\Delta_{\pom}$ values obtained in GLMM and KMR 
induce stronger screenings. Coupled to it 
there is an energy dependent renormalization
of $\Delta_{\pom}$ due to Pomeron loops corrections to the input Pomeron
propagator. The net result is that the effective $\Delta_{\pom}$ 
is reduced with energy. Along side this Phenomenon we expect $g_{3P}$, 
the triple Pomeron coupling, to monotonically reduce due to the 
continued reduction of the corresponding survival probability. 
\section{Data Analysis and Predictions}
\par
GLMM data analysis aims to fit the 
soft total, elastic, SD and DD cross sections and their  
forward slopes. Our output differential elastic cross sections 
and SD mass distributions are checked against the data.   
To this end we have compiled a comprehensive ISR-Tevatron data base.  
The conceptual approach of KMR is completely different. 
Their data base contains only the measured values of 
$d\sigma_{el}/dt$, which enables a prediction of $\sigma_{tot}$, 
and $d\sigma_{sd}/dtd(M^2/s)$.  
\par
In my opinion the KMR data base does not enable a decisive data analysis 
since their very limited data base is not sufficient to constrain 
their parameters. Specifically:
\newline
1) The features of the calculated $d\sigma_{el}/dt$ are, 
to a considerable extent, decoupled from the proposed dynamics 
and reflect the qualities of the b-profiles. 
Three KMR models\cite{RMK,LKMR,KMR}, with different dynamics and 
key parameters, reproduce, almost identically, 
the experimental $t$ distributions. 
The same is observed in GLMM where we obtain almost the same  
$d\sigma_{el}/dt$ output in a GW and a GW+${\pom}$ enhancement models. 
The output of LKMR is also supportive of this conclusion. 
\newline
2) The extensive LKMR analysis of CDF $d\sigma_{sd}/dtd(M^2/s)$  
indicates the importance of  
triple couplings with secondary Regge contributions.
More over, fitting this data requires a relative rescale of 
25$\%$ between the 540 and 1800 GeV normalization.
Consequently, KMR, who neglect  
$\sigma_{sd}$ and $\sigma_{dd}$, can not substantiate the need for  
their multi Pomeron sector just by reconstructing 
$d\sigma_{sd}/dtd(M^2/s)$. 
GLMM fit of $\sigma_{sd}$ and $\sigma_{dd}$ 
is supportive of Pomeron enhancement. 
LKMR analysis (which does not include $\sigma_{dd}$) 
suggests that the zero order Pomeron enhancement 
is sufficient to describe the available data.   
\newline
3) KMR neglect of quantitative assessment of their output 
adds to the ambiguity of their results. 
\begin{table}
\begin{tabular}{|l|l|l|l|}
\hline
& \,\,\,\,\,\,\,\,\,\,\,\,\,\,\,\,Tevatron
& \,\,\,\,\,\,\,\,\,\,\,\,\,\,\,\,\,\,\, LHC
& \,\,\,\,\,\,\,\,\, W=$10^5$ GeV  \\
& GLMM\,\,\,\,\,\,\,\,\,\,\,\,\,\,\,\,\,KMR
& GLMM\,\,\,\,\,\,\,\,\,\,\,\,\,\,\,\,\,KMR
& GLMM\,\,\,\,\,\,\,\,\,\,\,\,\,\,\,\,\,\,KMR \\
\hline
$\sigma_{tot}$( mb )
& 73.3 \,\,\,\,\,\,\,\,\,\,\,\,\,\,\,\,\,\,\,\,\,\,\,\,74.0
& 92.1\,\,\,\,\,\,\,\,\,\,\,\,\,\,\,\,\,\,\,\,\,\,\,\,\,\,88.0
&108.0\,\,\,\,\,\,\,\,\,\,\,\,\,\,\,\,\,\,\,\,\,\,\,\,98.0 \\
\hline
$\sigma_{el}$(mb)
& 16.3\,\,\,\,\,\,\,\,\,\,\,\,\,\,\,\,\,\,\,\,\,\,\,\,\,\,16.3
& 20.9\,\,\,\,\,\,\,\,\,\,\,\,\,\,\,\,\,\,\,\,\,\,\,\,\,\,20.1
& \,\,\,24.0\,\,\,\,\,\,\,\,\,\,\,\,\,\,\,\,\,\,\,\,\,\,\,\,22.9 \\
\hline
$\sigma_{sd}$(mb)
& \,\,\,9.8 \,\,\,\,\,\,\,\,\,\,\,\,\,\,\,\,\,\,\,\,\,\,\,\,\,10.9
& 11.8 \,\,\,\,\,\,\,\,\,\,\,\,\,\,\,\,\,\,\,\,\,\,\,\,13.3
& \,\,\,14.4  \,\,\,\,\,\,\,\,\,\,\,\,\,\,\,\,\,\,\,\,\,\,\,15.7 \\
$\sigma^{\mbox{low M}}_{sd}$
& \,\,\,8.6 \,\, \,\,\,\,\,\,\,\,\,\,\,\,\,\,\,\,\,\,\,\,\,\,\,\,4.4
& 10.5 \,\,\,\, \,\,\,\,\,\,\,\,\,\,\,\,\,\,\,\,\,\,\,\,\,\,5.1
& \,\,\,12.2\,\,\,\, \,\,\,\,\,\,\,\,\,\,\,\,\,\,\,\,\,\,\,\,\,\,5.7 \\
$\sigma^{\mbox{high  M}}_{sd}$
& \,\,\,1.2\,\,\,\,\,\,\,\,\,\,\,\,\,\,\,\,\,\,\,\,\,\,\,\,\,\,\,\,\,6.5
& \,\,\,1.3 \,\,\,\,\,\,\,\,\,\,\,\,\,\,\,\,\,\,\,\,\,\,\,\,\,\,\,\,8.2
& \,\,\,\,\,\,2.2\,\,\,\,\,\,\,\,\,\,\,\,\,\,\,\,\,\,\,\,\,\,\,\,10.0 \\
\hline
$\sigma_{dd}$(mb) & \,\,\,5.4
\,\,\,\,\,\,\,\,\,\,\,\,\,\,\,\,\,\,\,\,\,\,\,\,\,\,\,\,7.2
& \,\,\,6.1 \,\,\,\,\,\,\,\,\,\,\,\,\,\,\,\,\,\,\,\,\,\,\,\,\,13.4
& \,\,\,\,\,\,6.3\,\,\,\,\,\,\,\,\,\,\,\,\,\,\,\,\,\,\,\,\,\,\,\,17.3 \\
\hline
$\Lb\sigma_{el} + \sigma_{diff}\Rb/\sigma_{tot}$
& \,\,\,0.43\,\,\,\,\,\,\,\,\,\,\,\,\,\,\,\,\,\,\,\,\,\,\,\,\,\,0.46
& \,\,\,0.42\,\,\,\,\,\,\,\,\,\,\,\,\,\,\,\,\,\,\,\,\,\,\,\,\,\,0.53
& \,\,\,\,\,\,0.41\,\,\,\,\,\,\,\,\,\,\,\,\,\,\,\,\,\,\,\,\,\,\,\,0.57\\
\hline
\end{tabular}
\caption{Comparison of GLMM and KMR cross section outputs.}
\label{T}
\end{table}
\par
GLMM and KMR high energy Tevatron, LHC and Cosmic Rays
cross section predictions are summarized in Table 1.
The elastic and total cross section outputs of the two models are compatible
and, above the Tevatron, significantly lower than those 
obtained in models with no multi Pomeron contributions 
as a consequence of $\Delta_{\pom}$ renormalization.
GLMM and KMR predicted $\sigma_{sd}$ are compatible. However, KMR estimates 
of low mass diffraction are systematically lower, 
and for high mass systematically higher, than GLMM.
KMR define GW low mass diffraction when its rapidity
$Y \leq 3$. Multi Pomeron diagrams, which lead to high mass diffraction, are 
summed when Y > 3. Consequently, there is no overlap between the  
GW low mass and enhanced Pomeron high mass contributions.
In GLMM there is no Y cut of the GW contribution and the multi Pomeron diagrams 
are summed differently. As a result, the two contributions overlap. 
KMR $\sigma_{dd}$ is significantly larger than GLMM (see Table 1).
Alan Martin's talk at this meeting neglects the issue of double diffraction 
and I am not clear if KMR retract now from their excessive DD predictions. 
\section{Discussion}
\par
This presentation leads to an important technical conclusion that 
a fitted data analysis based on a 
diversified data base is essential when trying to establish the role 
of a new dynamical mechanism, multi Pomeron interactions in our context.
At present, the support for multi Pomeron enhancement 
beyond the zero order is marginal. 
The forthcoming first generation LHC and AUGER results
should provide crucial information on both fundamental and
current important problems discussed in this presentation.
\newline
1) Total and elastic cross sections calculated without the contribution
of multi Pomeron interactions are considerably larger
than those in which Pomeron enhancement is included. 
The calculated differences are model dependent.
This measurement has fundamental implication 
for the high energy limit of soft scattering.
\newline
2) A measurement of the diffractive cross sections should settle the model
differences between GLMM and KMR.
\newline
3) LHC measurements should enable us to estimate the rate at which the 
elastic scattering amplitude 
approaches the unitarity black disc bound.
$a_{el}(s,b)$ reaches this bound at a given $s$ and $b$ when, and only when, 
$A^S_{1,1}(s,b)$=$A^S_{1,2}(s,b)$=$A^S_{2,2}(s,b)$=1. 
Given GLMM adjusted parameters 
and the effect of $\Delta_{\pom}$ being reduced, the approach of $a_{el}(s,b=0)$ 
toward the black bound is very slow, well above LHC. 
\newline
4) Unitarity implied phenomena  
become more extreme at ultra high energies, 
as $a_{el}$ gets darker. 
Coupled to this, the inelastic diffractive 
cross sections becomes more and more 
peripheral and relatively smaller when compared to 
the elastic cross section. This is implied by the Pumplin bound applied to 
the GW low mass diffraction and the decrease of the survival probability 
applied to high mass diffraction. 
At the extreme, when $a_{el}(s,b)$ = 1, $a_{sd}(s,b)\,=\,a_{dd}(s,b)\,=\,0$. 
\newline
\newline
{\bf {Acknowledgments:}} 
This research was supported
in part by BSF grant $\#$ 20004019 and by
a grant from Israel Ministry of Science, Culture and Sport and
the Foundation for Basic Research of the Russian Federation.




\begin{thebibliography}{99}
\vskip0.5cm
\bibitem{heralhc}
E. Gotsman, E. Levin, U. Maor, E. Naftali and A. Prygarin:
{\it "HERA and the LHC-A workshop on the implications of
HERA for LHC physics: Proceedings Part A"} (2005) 221, {arXiv:hep-ph0601012}.
\bibitem{GLMM}
E. Gotsman, E. Levin, U. Maor and J.S. Miller: 
{arXiv:0805.2799[hep-ph]} (to be published in {\it Eur. Phys. J.}).
Denoted GLM.
\bibitem{RMK} 
M.G. Ryskin, A.D. Martin and V.A. Khoze:
{\it Eur. Phys. J.} {\bf C54} (2008) 199.
Denoted KMR.
\bibitem{LKMR}
E.G. Luna, V.A. Khoze, A.D. Martin and M.G. Ryskin:
{arXiv:0807.4115[hep-ph]}.
Denoted LKMR.
\bibitem{TT}
S.M. Troshin and N.E. Tyurin:
{\it Eur. Phys. J.} {\bf C39} (2005) 435. 
\bibitem{GW}
M.L. Good and W.D. Walker:
{\it Phys. Rev.} {\bf 120} (1960) 1857. Denoted GW.
\bibitem{KMR}
V.A. Khoze, A.D. Martin and M.G. Ryskin:
{\it Eur. Phys. J.} {\bf C18} (2000) 167. 

\end{thebibliography}
\end{document}